\documentstyle[12pt]{article}
\begin{document}
\title{{\bf CONSCIOUSNESS AND THE QUANTUM}
\thanks{Alberta-Thy-03-11, arXiv:yymm.nnnn [hep-th];
invited contribution for a special edition
of the Journal of Consciousness,
{\em Consciousness in the Universe},
edited by Roger Penrose and Stuart Hameroff
}}
\author{
Don N. Page
\thanks{Internet address:
profdonpage@gmail.com}
\\
Theoretical Physics Institute\\
Department of Physics, University of Alberta\\
Room 238 CEB, 11322 -- 89 Avenue\\
Edmonton, Alberta, Canada T6G 2G7
}
\date{(2011 Feb. 25)}

\maketitle
\large
\begin{abstract}
\baselineskip 19 pt

Sensible Quantum Mechanics or Mindless Sensationalism is a framework for
relating consciousness to a quantum universe.  It states that each
conscious perception has a measure that is given by the expectation
value of a corresponding quantum ``awareness operator'' in a fixed
quantum state of the universe.  The measures can be interpreted as
frequency-type probabilities for a large set of perceptions that all
actually exist with varying degrees of reality, so detailed theories
within this framework are testable.  The measures are not propensities
for potentialities to be actualized, so there is nothing indeterministic
in this framework, and no free will in the incompatibilistic sense.  As
conscious perceptions are determined by the awareness operators and the
quantum state, they are epiphenomena.  No fundamental relation is
postulated between different perceptions (each being the entirety of a
single conscious experience and thus not in direct contact with any
other), so SQM or MS, a variant of Everett's ``many-worlds'' framework,
is a ``many-perceptions'' framework but not a ``many-minds'' framework.

\end{abstract}

\normalsize

\baselineskip 18.2 pt

\newpage

I am a physicist, and we physicists are trying to find and understand at
least one theory that will give a good description and explanation
of our universe.  It is typically preferred that such a theory have
simple principles, an elegant form, and yet make precise statements. 
For these purposes, mathematical theories are often the ideal.

On the other hand, physics is usually regarded as necessarily resting
upon observations (in contrast to, say, pure mathematics, which in
principle can be divorced from experience, though in practice it too is
usually based on observed patterns).  However, observations themselves
seem less simple, elegant, and precise than what physicists would often
want for their theories.  I am of the opinion that this is one of the
causes for the tendency to regard the simple, elegant, and precise
elements of theories of physics as more real than the apparently
complex, sometimes ugly, and usually imprecise observations that we pay
lip service to as the foundation of physics.  Whatever we cannot
understand in terms of the simple, elegant, precise elements of our
mathematical theories, we tend to dismiss as less real.

In particular, I personally regard my own first-person subjective
experience of consciousness as overwhelming evidence of its existence,
but its apparent complexity, inelegance, and imprecision often seems to
lead many physicists to dismiss it as less real than, say, elementary
particles, spacetime, or quantum states.  I am not at all decrying
simplicity, elegance, and precision, which I do take to be extremely
important, but I also do not wish to dismiss such a central feature of
our experience as consciousness.

In fact, I take the observations that are considered to be the
foundation of physics to be most simply and fundamentally conscious
perceptions.  This viewpoint raises the question of how consciousness is
related to the structures that are more common in current theories of
physics.  In particular, the most fundamental structure of many of the
current theories of physics is the quantum framework.  Although we
cannot be certain that our universe really is quantum, such a hypothesis
helps explain so much of both our present theories and our observations
that I shall take it as one of my central working hypotheses, along with
the hypothesis of the existence of conscious perceptions.

In view of our desire to formulate the most simple, elegant, precise
theories possible, I am not be content with leaving these two
fundamental hypotheses unrelated but instead want to integrate them
together.  How might that be done?

Here I wish to describe and explore a framework I have developed for
relating consciousness and the quantum, which I have called Sensible
Quantum Mechanics (Page, 1995) or Mindless Sensationalism (Page, 2003). 
I should emphasize that this is so far just a framework, since the
details to make it into a proper precise theory are not yet known.  But
even just the framework itself has various consequences that may be
explored.

Because Sensible Quantum Mechanics builds upon quantum theory, I should
first say what I regard that pillar to be.  I regard the essence of
quantum theory to be a $C^*$-algebra of quantum operators and a quantum
state that gives expectation values to each operator.  

I cannot go into precise details here, but let me give a crude
dscription of these elements.  Quantum operators are mathematical
entities that may be adjointed, multiplied by complex numbers, added
together, or multiplied together to give other operators.  For example,
if $A$ and $B$ are two operators and $c$ is a complex number (so $c =
a+ib$ with $a$ and $b$ real numbers and $i$ the square root of $-1$),
then the adjoints $A^\dag$ and $B^\dag$, $cA$, $cB$, $A+B$, $AB$, and
$BA$ are also quantum operators.  (It is not assumed that $AB = BA$; the
order of two operators that are multiplied together generally matters.) 
An operator $A$ might represent the position of a particle, and $B$
might represent its momentum, but for the general structure we do not
need to assign specific meanings to the operators.

A quantum state may be regarded as a positive linear functional $\sigma$
on the quantum operators, a rule for assigning a complex number to each
quantum operator that is called its expectation value in that quantum
state.  For example, the expectation value of the operator $A$ in the
quantum state $\sigma$, denoted $\sigma[A]$, represents a particular
complex number associated with that operator.  This rule is required to
have the form that $\sigma[cA] = c\sigma[A]$ for any operator $A$ and
complex number $c$, so the expectation value of the operator $cA$ that
is the complex number $c$ multiplied by the operator $A$ is simply $c$
times the expectation value of $A$.  The rule is also required to give
$\sigma[A+B]=\sigma[A]+\sigma[B]$ and $\sigma[I] = 1$, where $I$ is the
identity operator such that $IA = AI = A$ for each quantum operator $A$,
as well as to make $\sigma[A^\dag A]$ a nonnegative real number.

Quantum theory is often regarded as having other basic elements, but I
shall regard them as either being part of this basic formalism or as not
really being a necessary part of quantum theory.  For example, often one
talks about the time evolution of a quantum state, but one can
reformulate this into the Heisenberg picture in which the quantum state
stays fixed and the operators change with time, and then one can
re-interpret the time dependence of each operator as representing a
whole family of operators, each labeled by a time parameter in addition
to other labels of what the operators are.  The dynamics of the
operators in the usual approach would expressed in terms of the algebra
of all the operators in all the families labeled by the time parameter
and by the other labels.  In this view, there would be nothing
fundamentally special about time; it would just be one of many labels
for the operators.

The other pillar of Sensible Quantum Mechanics is consciousness.  I
shall assume here that there is a countable discrete set $M$ of all
possible conscious experiences or perceptions $p$.  By a ``conscious
experience,'' I mean all that one is consciously aware of or consciously
experiencing at once.  Lockwood (2003) has called this a ``phenomenal
perspective'' or ``maximal experience'' or ``conscious state.'' It could
also be expressed as a total ``raw feel'' that one has at once.

Because I regard the basic conscious entities to be the conscious
experiences themselves, which might crudely be called sensations if one
does not restrict the meaning of this word to be the conscious responses
only to external stimuli, and because I doubt that these conscious
experiences are arranged in any strictly defined sequences that one
might define to be minds if they did exist, my framework has sensations
without minds and hence may be labeled Mindless Sensationalism (Page,
2003).  In this way the framework of Mindless Sensationalism proposed
here is a particular manifestation of Hume's ideas (Hume, 1988), that
``what we call a {\it mind}, is nothing but a heap or collection of
different perceptions, united together by certain relations, and
suppos'd, tho' falsely, to be endow'd with a perfect simplicity and
identity'' (p. 207), and that the self is ``nothing but a bundle or
collection of different perceptions'' (p. 252).  As he explains in the
Appendix (p. 634), ``When I turn my reflexion on {\it myself}, I never
can perceive this {\it self} without some one or more perceptions; nor
can I ever perceive any thing but the perceptions.  'Tis the composition
of these, therefore, which forms the self.''  (Here I should note that
what Hume calls a perception may be only one {\it component} of the
``phenomenal perspective'' or ``maximal experience'' (Lockwood, 1989)
that I have been calling a perception or conscious experience $p$, so
one of my $p$'s can include ``one or more perceptions'' in Hume's
sense.)

I should also emphasize that by a conscious experience, I mean the
phenomenal, first-person, ``internal'' subjective experience, and not
the unconscious ``external'' physical processes in the brain that
accompany these subjective phenomena.  In his first chapter, Chalmers
(1996) gives an excellent discussion of the distinction between the
former, which he calls the phenomenal concept of mind, and the latter,
which he calls the psychological concept of mind. In his language, what
I mean by a conscious experience (and by other approximate synonyms that
I might use, such as perception or sensation or awareness) is the
phenomenal concept, and not the psychological one.

The next idea is that not all possible conscious perceptions $p$ occur
equally, but that there is a normalized measure $w(p)$ associated with
each one (a nonnegative real number which sums to unity when one adds up
the values for all the $p$'s in the full set $M$).  This measure in some
sense gives the level or degree of reality that the conscious perception
$p$ has.  Perceptions with large measures have high degrees of reality,
whereas perceptions with very low measures have tiny degrees of reality
and effectively can be ignored.  One can also say that the weight $w(p)$
is analogous to the probability for the conscious experience $p$, but it
is not to be interpreted as the probability for the bare {\it existence}
of $p$, since any conscious experience $p$ exists (is actually
experienced) if its weight is positive, $w(p) > 0$.  Rather, $w(p)$ is
to be interpreted as being proportional to the probability of {\it
getting} this particular experience if a random selection were made.

Because the specification of the conscious experience $p$ completely
determines its content and how it is experienced (how it feels), the
weight $w(p)$ has absolutely no effect on that---there is absolutely no
way within the experience to sense anything directly of what the weight
is.  A toothache within a particular conscious experience $p$ is
precisely as painful an experience no matter what $w(p)$ is.  It is just
that an experience with a greater $w(p)$ has a greater degree of
existence and is more likely in the sense of being more probably chosen
by a random selection using the weights $w(p)$.

Finally, Sensible Quantum Mechanics assumes the connection between
consciousness and the quantum in the form that for each conscious
perception $p$, there is an associated quantum ``awareness operator''
$A(p)$, and that the measure for the conscious perception $p$ is the
expectation value, in the quantum state of the universe, of the
corresponding experience operator, $w(p) = \sigma[E(p)]$.

One can summarize this by saying that Sensible Quantum Mechanics or
Mindless Sensationalism is given by the following three basic postulates
or axioms (Page, 1995, 2003):

 {\bf Quantum World Axiom}:  The ``quantum world'' $Q$ is completely
described by an appropriate $C^*$-algebra of operators $O$ and by a
suitable state $\sigma$ (a positive linear functional of the operators)
giving the expectation value $\sigma[O]$ of each operator $O$.

 {\bf Conscious World Axiom}:  The ``conscious world'' $M$, the
countable discrete set of all conscious experiences or perceptions $p$,
has a fundamental normalized measure $w(p)$ for each perception $p$.

 {\bf Psycho-Physical Parallelism Axiom}: The measure $w(p)$ for each
conscious experience $p$ is given by the expectation value of a
corresponding quantum ``awareness operator'' $A(p)$ in the state
$\sigma$ of the quantum world, $w(p) = \sigma[A(p)]$.

One might note that in comparison with the more general assumptions of
(Page, 1995, 2003), here for simplicity I am making the more restrictive
hypothesis that the set $M$ of all conscious perceptions $p$ is a
countable discrete set.  I am also assuming that the measure is
normalized, $\sum_p w(p) = 1$.

The Psycho-Physical Parallelism Axiom is the simplest way I know to
connect the quantum world with the conscious world. One could easily
imagine more complicated connections, such as having $w(p)$ be a
nonlinear function of the expectation values, say $m(p)$, of a positive
``experience operator'' $E(p)$ depending on the $p$ (Page, 1995, 2003).
Instead, my Psycho-Physical Parallelism Axiom restricts the function to
be linear in the expectation values. In short, I am proposing that the
psycho-physical parallelism is {\it linear}.

Of course, the Psycho-Physical Parallelism Axiom, like the Quantum World
Axiom, is here also deliberately vague as to the form of the awareness
operators $A(p)$, because I do not have a detailed theory of
consciousness, but only a framework for fitting it with quantum theory.
My suggestion is that a theory of consciousness that is not inconsistent
with bare quantum theory should be formulated within this framework. I
am also suspicious of any present detailed theory that purports to say
precisely under what conditions in the quantum world consciousness
occurs, since it seems that we simply don't know yet. I feel that
present detailed theories may be analogous to the cargo cults of the
South Pacific after World War II, in which an incorrect theory was
adopted, that aircraft with goods would land simply if airfields and
towers were built.

Since all conscious perceptions $p$ with $w(p) > 0$ really occur in the
framework of Sensible Quantum Mechanics or Mindless Sensationalism, it
is completely deterministic if the quantum state and the awareness
operators $A(p)$ are determined:  there are no random or truly
probabilistic elements in SQM or MS.  Neither is there any free will in
the incompatibilist sense, and consciousness may be viewed as an
epiphenomenon (Page, 1995, 2003).  Nevertheless, because the framework
has normalized measures $w(p)$ for conscious perceptions, these can be
interpreted as probabilities for the perceptions, given the theory.  In
particular, one can interpret the measure $w_i(p_j)$ that a theory $T_i$
assigns to one's particular perception $p_j$ as the likelihood of the
theory.  Then if one assigns different SQM or MS theories prior
probabilities $P(T_i)$, one can use Bayes' theorem to calculate the
posterior probability of the theory, given the observation or conscious
perception $p_j$, as $P(T_i|p_j) = P(T_i)w_i(p_j)/\sum_k
P(T_k)w_k(p_j)$.  In this way different SQM or MS theories are testable.

A major problem at the frontier of theoretical cosmology is essentially
to develop one or more theories that give the measures $w(p)$ for
conscious perceptions, except that most theorists are hesitant to focus
on conscious perceptions and hence ask for the probability of an
observation $O_j$ given a theory $T_i$, $P(O_j|T_i)$.  It is usually
left rather vague what is supposed to constitute an observation.  For me
the most fundamental entities that can be identified with observations
are conscious perceptions, so I would take $P(O_j|T_i)$ to be
$w_i(p_j)$, the normalized measure that theory $T_i$ assigns to the
conscious perception $p_j$.  In Sensible Quantum Mechanics, a theory
$T_i$ would assign an awareness operator $A_j = A(p_j)$ to each
conscious perception and give a quantum state $\sigma_i$ so that
$P(O_j|T_i) = w_i(p_j) = \sigma_i[A_j]$, the expectation value of the
awareness operator in the quantum state.  (Here, for compactness, I do
not explicitly display the dependence of the $A_j$ operators on the
theory $T_i$, but different theories can differ not only in their
quantum states but also in their awareness operators.)

For theorists hesitant to identify observations with conscious
perceptions, they may still wish to say that the probability of an
observation $O_j$ given a theory $T_i$ is $P(O_j|T_i) = \sigma_i[A_j]$.
In this generalized view, $A_j$ is simply the operator in the theory
$T_i$ whose expectation value in the quantum state given by that theory
gives the probability of the observation $O_j$.

Quantum theories of this generalized form (whether or not an observation
is taken to be a conscious perception) can be considered to have three
parts:  (1) the algebra of the full set of quantum operators, (2) the
quantum state $\sigma_i$, and (3) the particular operators $A_j$ for
each observation (or for each conscious perception in Sensible Quantum
Mechanics or Mindless Sensationalism).  Part (1) includes the dynamical
laws of physics, which historically have often been na\"{\i}vely called
a `Theory of Everything' or TOE, though it certainly is not.  Part (2)
includes the boundary conditions that specify which solution of the
dynamical laws describes our actual universe, but even Part (1)
augmented with Part (2) is not sufficient.  Part (3) includes the rules
for extracting the probabilities of observations from the quantum state.

The logical independence of Part (3) is becoming widely recognized with
the {\it measure problem} of cosmology (Linde, 1986; Garcia-Bellido et
al., 1994; Vilenkin, 1995a,b; Guth, 2000; Tegmark 2005; Aguirre, 2007;
De Simone et al., 2008; Linde \& Noorbala, 2010; Bousso et al., 2010). 
If a universe had a definite number $N_j$ of occurrences (each
occurrence with the same degree of reality) of each kind of observation
$O_j$, and a finite total number of occurrences $N = \sum_k N_k$, it
would be natural to say that the probability of the observation $O_j$ is
the fraction of the number of its occurrences, $P(O_j) = N_j/N$. 
However, theories of eternal inflation (Linde, 1986; Garcia-Bellido et
al., 1994; Vilenkin, 1995a,b; Guth, 2000; Tegmark 2005; Aguirre, 2007;
De Simone et al., 2008; Linde \& Noorbala, 2010; Bousso et al., 2010)
suggest that the universe may have expanded to become infinitely large,
in which case most of the numbers $N_j$ of occurrences are infinite, and
the ratio $N_j/N$ is ambiguous.  Therefore, a lot of work has gone into
different proposed schemes for regularizing the infinities.

I have pointed out that even in finite universes, quantum uncertainties
in the numbers of occurrences also leads to ambiguities in the
probabilities of observations (Page, 2008, 2009a,b,c, 2010).  In
particular, I have shown that Born's rule does not work in the sense
that the operators $A_j$ cannot be projection operators, so that one
must choose other operators, and the ambiguity of that choice is the
measure problem.  The ambiguity occurs even for finite universes, but it
is particularly severe for infinite universes.  So whether or not the
operators $A_j$ whose expectation values give the the probabilities of
observations are interpreted to be awareness operators in Sensible
Quantum Mechanics or Mindless Sensationalism (in which the observations
are conscious perceptions), it is now recognized that there is the
challenge of finding these operators, in addition to the challenge of
finding the dynamics or algebra of all operators and the quantum state.

In conclusion, I am proposing that Sensible Quantum Mechanics or
Mindless Sensationalism is the best framework we have for understanding
the connection between consciousness and the quantum universe.  Of
course, the framework would only become a complete theory once one had
the set of all conscious experiences, the awareness operators, and the
quantum state of the universe.

This research was supported in part by the Natural Sciences and
Engineering Research Council of Canada.


\baselineskip 23pt

\section*{Bibliography}

Aguirre, A.~(2007) On making predictions in a multiverse: Conundrums,
dangers, and coincidences.  In: Carr, B.~J.~(Ed.), Universe or
Multiverse?, Cambridge University Press,
Cambridge, UK, pp.\ 367-386 [arXiv:astro-ph/0506519].

Bousso, R.,Freivogel, B.,Leichenauer, S., Rosenhaus, V. (2010). 
Geometric origin of coincidences and hierarchies in the landscape. 
arXiv:1012.2869 [hep-th].

Chalmers, D.~J.~(1996).  The Conscious Mind:  In Search of a Fundamental
Theory.  Oxford University Press, New York, USA.

Garcia-Bellido, J., Linde, A.~D., Linde, D.~A.~(1994).  Fluctuations of
the gravitational constant in the inflationary Brans-Dicke cosmology. 
Physical Review D, 50, 730-750 [arXiv:astro-ph/9312039].

Guth, A.~H.~(2000).  Inflation and eternal inflation.  Physics Reports,
333, 555-574 (2000) [arXiv:astro-ph/0002156].

Hume, D.~(1988).  A Treatise of Human Nature.  Clarendon Press, Oxford,
UK.

Linde, A.~D.~(1986).  Eternally existing self-reproducing chaotic
inflationary universe.  Physics Letters B, 175, 395-400.

Linde, A., Noorbala, M. (2010).  Measure problem for eternal and
non-eternal inflation.  Journal of Cosmology and Astroparticle Physics,
1009, 008 [arXiv:1006.2170 [hep-th]].

Lockwood, M.~(1989).  Mind, Brain and the Quantum:  The Compound `I.' 
Basil Blackwell Press, Oxford, UK.

Lockwood, M.~(2003) Consciousness and the quantum world: Putting qualia
on the map. In Smith, Q., Jokic, A.(Eds.), Consciousness: New
Philosophical Perspectives, Oxford University Press, Oxford, pp.
447-467.

Page, D.~N.~(1995). Sensible quantum mechanics:  Are only perceptions
probabilistic?  arXiv:quant-ph/9506010.

Page, D.~N.~(2003) Mindless sensationalism:  A quantum framework for
consciousness. In Smith, Q., Jokic, A.(Eds.), Consciousness: New
Philosophical Perspectives, Oxford University Press, Oxford, pp.
468-506 [arXiv:quant-ph/0108039].

Page, D.~N.~(2008).  Cosmological measures without volume weighting. 
Journal of Cosmology and Astroparticle Physics, 0810, 025
[arXiv:0808.0351 [hep-th]].

Page, D.~N.~(2009a).  Insufficiency of the quantum state for deducing
observational probabilities.  Physics Letters B, 678, 41-44
[arXiv:0808.0722 [hep-th]].

Page, D.~N.~(2009b).  The Born rule fails in cosmology.  Journal of
Cosmology and Astroparticle Physics, 0907, 008 [arXiv:0903.4888
[hep-th]].

Page, D.~N.~(2009c).  Born again.  arXiv:0907.4152 [hep-th].

Page, D.~N.~(2010).  Born's rule is insufficient in a large universe. 
arXiv:1003.2419 [hep-th].

De Simone, A., Guth, A.~H.,Salem, M.~P., Vilenkin, A. (2008). 
Predicting the cosmological constant with the scale-factor cutoff
measure.  Physical Review D, 78, 063520 [arXiv:0805.2173 [hep-th]].

Tegmark, M.~(2005).  What does inflation really predict?  Journal of
Cosmology and Astroparticle Physics, 0504, 001 [arXiv:astro-ph/0410281].

Vilenkin, A.~(1995a).  Predictions from quantum cosmology.  Physical
Review Letters, 74, 846-849 [arXiv:gr-qc/9406010].

Vilenkin, A.~(1995b).  Making predictions in eternally inflating
universe.  Physical Review D, 52, 3365-3374 [arXiv:gr-qc/9505031].

\end{document}